\documentclass[reprint]{JASAnew}
\pdfoutput=1
\newcommand{\add}[1]{#1}
\newcommand{\delete}[1]{}
\usepackage{color}
\usepackage[caption=false]{subfig}
\usepackage{tikz}
\usetikzlibrary{chains,shapes,arrows,positioning,arrows.meta,patterns,calc}
\usepackage{pgfplots}
\pgfplotsset{compat=1.10}
\usepgfplotslibrary{fillbetween,smithchart,statistics}
\usepackage{standalone}
\usepackage{epstopdf}
\usepackage{algorithm,setspace}
\usepackage[noend]{algpseudocode}
\usepackage{xpatch}
\newcommand{\NORI}{{\textsc{Nori}}}
\usepackage{enumitem}
\newcolumntype{L}[1]{>{\raggedright\let\newline\\\arraybackslash\hspace{0pt}}m{#1}}
\newcolumntype{C}[1]{>{\centering\let\newline\\\arraybackslash\hspace{0pt}}m{#1}}
\newcolumntype{R}[1]{>{\raggedleft\let\newline\\\arraybackslash\hspace{0pt}}m{#1}}
\makeatletter
\xpatchcmd{\algorithmic}{\itemsep\z@}{\itemsep=2ex plus5pt}{}{}
\makeatother

\begin{document}
%% the square bracket argument will send term to running head in
%% preprint, or running foot in reprint style.

%  Beginning of title page for Preprint option --------------------------------- %
\title[]{Non-intrusive speech intelligibility prediction using automatic speech recognition derived measures}

% ie
%\title[JASA/Sample JASA Article]{Sample JASA Article}

\author{Mahdie Karbasi}%\email{mahdie.karbasi@rub.de}
\affiliation{Institute of Communication Acoustics, Faculty of Electrical Engineering and Information Technology,	Ruhr University Bochum, Universit{\"a}tsstr. 150, 44801 Bochum, Germany}
\author{Stefan Bleeck}			% \email{S.Bleeck@soton.ac.uk}
\affiliation{Institute of Sound and Vibration Research, University of Southampton, SO17 1BJ, UK}
\author{Dorothea Kolossa$^1$}			% \email{author.two@someplace.edu}

%\address{%Institute of Communication Acoustics, 
%	$^1$Cognitive Signal Processing Group,
%	Ruhr-Universit\"{a}t Bochum, 44801 Bochum, Germany \\
%	$^2$International Computer Science Institute, Berkeley, USA\\
%}

%\affiliation{ Cognitive Signal Processing Group, Ruhr-Universit\"{a}t Bochum, 44801 Bochum, Germany}
% ie
%\author{Author One}
%\author{Author Two}
%\author{Author Three}

% ie
%\affiliation{Department1,  University1, City, State ZipCode, Country}

%% for corresponding author
%\email{}
%% for additional information
%\thanks{}

% ie
% \author{Author Four}
% \email{author.four@university.edu}
% \thanks{Also at Another University, City, State ZipCode, Country.}

%% For preprint only,
%  optional, if you want want this message to appear in upper left corner of title page
% \preprint{}

%ie
%\preprint{Author, JASA}		

% optional, if desired:
%\date{\today} 

%============================================================================= 
\begin{abstract}
%\ifversionB
The estimation of speech intelligibility is still far from being a solved problem. Especially one aspect is problematic: most of the standard models require a clean reference signal in order to estimate intelligibility. \delete{This is a problem in two respects. A reference signal is often unavailable in practice. Also, comparing the signal with a reference leads to unrealistic results when speech signals processing is carried out with the help of a recognize\&synthesize approach, where the voice of the speaker is replaced by that of a synthesizer.} \add{This is an issue of some significance, as a reference signal is often unavailable in practice.} 
In this work, therefore a \delete{fully reference-free} \add{non-intrusive} speech intelligibility estimation framework is presented. In it, human listeners' performance in keyword recognition tasks is predicted using intelligibility measures that are derived from models trained for automatic speech recognition (ASR). One such ASR-based and one signal-based measure are combined into a full framework, the proposed \emph{NO-Reference Intelligibility} (\NORI{}) estimator, which is evaluated in predicting the performance of both  normal-hearing and hearing-impaired listeners in multiple noise conditions. It is shown that the \NORI{} framework even outperforms the widely used reference-based \add{(or intrusive)} short-term objective intelligibility (STOI) measure in most considered scenarios, while being applicable in fully blind scenarios \add{with no reference signal or transcription}, creating perspectives for online and personalized optimization of speech enhancement systems.

\end{abstract}
%\begin{keywords}
%speech intelligibility prediction, non-intrusive measures, speech assessment, automatic speech recognizer
%\end{keywords}
%============================================================================= 
%% pacs numbers not used
\maketitle
%  End of title page for Preprint option --------------------------------- %

%% See preprint.tex/.pdf or reprint.tex/.pdf for many examples
%============================================================================= 
\section{Introduction}\label{sec:Intro} 
%Speech processing algorithms are used in many every-day communication applications such as medical devices like hearing aids or cochlear implants, mobile phones or other speech-based computer interfaces (smart hubs) that are becoming ever more prevalent in daily life.
%Moving forward, there will be an increased need for \delete{better} \add{artificially intelligent speech processing} in hearing aids, as in aging populations, more people will benefit from hearing aids for which machine-learning offers a multitude of notable improvements. Moreover, speech-based telecommunication systems are also becoming ever more widely used in daily life and intelligent machines like digital assistants increasingly use speech as a rich human-machine communication modality. 
%As a result, substantial effort is being devoted to develop \delete{better} \add{more accurate} and reliable measures for evaluating speech enhancement and reproduction, specifically speech intelligibility (SI).
%
%
%In general, SI is defined as the percentage of words or phonetic units that can be recognized by a human listener. To measure SI accurately, ultimately therefore subjective tests with human listeners must be conducted. But performing such tests is cumbersome, expensive and slow and it is also a burden for the participants. Thus, there is a strong motivation to develop objective methods to asses SI without much involvement of the human listener. 

The intelligibility of a speech signal, defined as the percentage of words or phonetic units that can be recognized by a human listener, is dependent on a large number of factors, including the loudness of the speech, ambient noise, characteristics of the transmission channel or speaking style~\cite{AI}. A complete description of all parameters is very complex, and although there have been many (simplified) efforts to predict speech intelligibility (SI) objectively, this has not been achieved with\delete{sufficient precision in all circumstances} \add{equally high accuracy over different conditions such as non-linearly distorted or reverberated speech. Also, modeling hearing-impaired listeners and predicting their performance is still a challenging task}. 

Speech intelligibility prediction methods can be divided into two categories: \emph{intrusive} methods (reference-based approaches) require the clean reference signal. \emph{Non-intrusive}  (reference-free methods) predict the intelligibility without the need for a clean reference \add{signal}. 

%These methods usually utilize comparison measures such as signal-to-noise ratio (SNR) or correlation coefficient between a representation of the distorted/processed signal and its corresponding clean signal in the time-frequency domain.
Early work on intelligibility prediction exclusively used intrusive methods, as this is conceptually much simpler. For instance, the articulation index (AI)~\cite{AI}, the speech intelligibility index (SII)~\cite{SII}, and the speech transmission index (STI)~\cite{STI} were the first methods developed. \delete{All these methods} Both AI and SII work by first estimating the SNRs within psychoacoustically motivated frequency bands, and then using the weighted sum of those estimates as a measure of the speech intelligibility. \add{The speech transmission index (STI) is calculated using the weighted average of reductions in temporal envelope modulation by estimating the modulation frequency function.
%More advanced methods were subsequently developed based on the same concept of estimating SNRs using speech signal representations in auditory models~\cite{snrLoss_Loizou, sEPSM-2017yamamo}. 
The speech-based envelope power spectrum model (sEPSM)~\cite{sEPSM2011} and later the multi resolution sEPSM (mr-sEMSP)~\cite{sEPSM-mr2013} were introduced to overcome the drawbacks of STI in predicting non-linearly distorted or reverberated speech.}
%\delete{Complex methods (like SII) additionally use observed nonlinearities in the perception process to improve on earlier methods. However, these simple methods only work in restricted acoustic scenarios, and none is suitable to predict hearing-impaired listeners' performance.}

An improved, and well-established \add{intrusive} measure for speech intelligibility prediction today is the short-time objective intelligibility (STOI) measure~\cite{STOI}, which has become a common benchmark in the field of speech processing~\cite{STOI_application1,STOI-binMask_ICassp2015}. STOI is based on the correlation between the test signal and the reference signal within one-third octave frequency bands over short time segments.
Regardless of the reported high correlation between STOI and human speech recognition performance in different acoustic scenarios, it does not perform well for distorted (processed) speech or reverberant acoustic conditions.
Such distortions are especially a problem when assessing signal processing algorithms that are used in hearing aids, as they introduce further non-linear distortions, e.g. by compressing the dynamic range, that lead to subsequently reduced speech intelligibility. 
\add{To overcome this, further studies have been conducted to improve the original STOI metric to work in a wider range of conditions. For instance, the Extended STOI (ESTOI)~\cite{ESTOI} was introduced to perform better in scenarios with fluctuating noise. ESTOI is based on energy-normalized short-time spectrograms that are orthogonally decomposed into subspaces which are important for intelligibility.}

While the correlation-based measures such as STOI are limited to second-order statistics, it has been demonstrated that higher-order statistics improve SI predictions by using the mutual information (MI) between the test signal and reference ~\cite{Taghia2014,MI_Tall}. 
\add{Speech signals are sparsely encoded in the time-frequency domain and in this representation, the regions with higher energy of speech relative to the masker are called \emph{glimpses}. It has been shown that the reliable glimpses, which have an adequate SNR, are contributing to the intelligibility of speech~\cite{glimpsing_cooke2006,li2012_glimpsing}  and can be used to predict the intelligibility instrumentally~\cite{glimpse-tang2014thesis,glimpseBinaural2016tang}. 

Such methods provide an estimate of the average intelligibility over the entire speech signal. Also, they usually require longer segments of speech in order to achieve more accurate predictions of intelligibility.}
Assessing the intelligibility of speech not at the signal level but at the phoneme level has also attracted some recent attention. \citet{PhonClassConditionalProb_ullmann2015} suggest to use the distance between the phoneme posterior probabilities of the distorted and the reference signal, which are estimated by a deep neural network (DNN).
\citet{posteriograms_2017} introduce the mean temporal distance between posteriograms as a predictor of human listening effort.
%Phoneme posteriorgrams are deployed in assessing the quality of speech signals in hearing aids equipped with spatial filtering~\cite{posteriograms_spille2016}. It is shown that the phoneme entropy, estimated by computing DNN-based posteriorgrams, is dependent on the sound source location and the steering direction of the beamformer in a hearing aid. In a following work~\cite{posteriograms_2017}, the mean temporal distance between posteriograms is introduced as a predictor of human listening effort.
%
%Neurograms, obtained from auditory nerve models, are another source of information that have been used by~\citet{Neurogram_SimilarityIndex_Hines2012} to predict the phoneme recognition accuracy. For this purpose, the neurograms of a reference and degraded speech signal are compared using a similarity index measure. 
%Mamun et al. have also proposed a neurogram-based intelligibility measure, which uses orthogonal polynomials for estimating the amount of recognizable information in speech~\cite{Neurogram_OrthogonalMeasure_Mamun2015}. They have shown that this approach can predict the intelligibility score across the entire range of hearing loss (normal-hearing to severely impaired).

All methods described so far are \textit{intrusive} or reference-signal-based.
%, i.e., they require the clean reference signal to estimate the SI. 
However, in many applications the clean signal is not available or not practical to access. It is therefore important to also investigate non-intrusive methods with the goal of predicting the intelligibility of speech without the need for a reference signal. 
In recent work, \citet{FADE-schadler2015} have chosen a non-intrusive approach to develop the \emph{framework for auditory discrimination experiments} (FADE), which predicts speech reception thresholds (SRTs) for a German matrix sentence test.
%under different stationary and fluctuating noise conditions.
In FADE, reference HMMs are trained on specific target speech tokens as well as noise types that are used in the testing phase. %\add{This model is dependent on the statistical reference models, which are trained on specific target speech tokens as well as noise types}.
To individualize FADE for hearing-impaired listeners, audiogram thresholds and supra-threshold characteristics, \add{(distortions),} are used to estimate the spectrogram that is employed during the process of the speech feature extraction~\cite{FADE_modelingHI_ref}.  
\add{Another ASR-based framework has been introduced by~\citet{spille2018DNN} to predict SRTs in different noisy scenarios. They use a hybrid DNN/HMM structure to identify words from a German matrix sentence test. Phone-based HMMs with context-dependent triphone models are used for training. 
%It is shown that the DNN-based ASR predictions outperform intrusive SI measures such as STOI and mr-sEPSM in the considered scenarios with speech shaped noise or speech like maskers.
}
%\add{Two methods of non-intrusive SI estimation have been described that are based on automatic speech recognition: a) using task-specific statistical reference models~\cite{FADE-schadler2015},  and b) using models that have not been trained specifically for the input test signal~\cite{spille2018DNN}.}
%The underlying approach in both methods is to first learn the distribution of human recognition scores using degraded speech, and then to predict the intelligibility of unseen data using statistical models.%~\cite{Nemala2010,Sharma_LCIA,Sharma_LCIAassesment}
%Both approaches vary in the way that features are extracted and in the underlying model, with generative or discriminative models as the principal options.
In \add{addition to statistical models, in} another approach~\cite{SIP_CNN}, a convolutional neural network has been used to learn and predict the speech intelligibility non-intrusively. 

\delete{In addition to speech intelligibility, there is an equivalent drive to develop better objective models of speech quality. This is equally important for example when developing hearing aids, as it has been shown that listeners only accept a limited degradation of speech quality for the benefit of improved intelligibility \cite{bleeck2015_speechQuality}. Equivalent to SI, the most accurate method for evaluating speech quality is through costly and time consuming subjective listening tests. Often used objective measures for speech intelligibility include frequency-weighted segmental SNR (fwsegSNR) and the Perceptual Evaluation of Speech Quality (PESQ) measure, which have both been shown to predict perceived speech quality for normal-hearing listeners well \cite{evalQMeasures_loizou}. Another objective measure, the Hearing Aid Speech Quality Index (HASQI) was introduced to predict speech quality specifically for hearing aid users \cite{HAAQI}.} 

One issue that none of the described methods is conceptually able to model is the question of prior knowledge. Humans can of course use experience, knowledge about context and prior knowledge about the characteristics of speech units (such as phonemes) when listening to speech~\cite{fingscheidt2013phonetic}. 
Therefore a complete objective model for SI and speech quality must also take the phonetic contextual information into account. Otherwise, for example comparing the processed speech only to a signal-based reference might lead to unreasonably low intelligibility estimates in scenarios where the speech is strongly modified or even partially synthesized in the enhancement stage.
With the goal of making similar knowledge available to SI measures, a non-intrusive method has been introduced by~\citet{THMM-STOI,Blind_THMM-STOI} that uses the distorted input speech and its correspondent transcription to synthesize an estimated reference. In this method, the estimated reference signal can be used in intrusive measures (such as STOI) to predict intelligibility.

In addition to the above approaches to provide an estimate of the \emph{average} intelligibility over the entire speech signal, there are also so-called \textit{microscopic} SI prediction models. These predict the intelligibility of smaller units of speech, such as words or phonemes, taking knowledge about the functioning of the human auditory process into account~\cite{jurgens2009microscopic,Juergens2010ChallengingTS}. This allows for individual tailoring of the process, for example by taking wider filter shapes or higher thresholds into account.
Microscopic methods promise to be more precise than macroscopic models in predicting intelligibility and in diagnosing problems due to specific phoneme confusions. However, they still need further improvements to become more robust against variations of the test scenarios in future work. 

In order to utilize the advantage of microscopic methods \add{in predicting the intelligibility of small units of speech accurately} and inspired by previous work on non-intrusive methods, in this paper we investigate the possibility of using automatic speech recognition to extract non-intrusive measures to predict SI. As a secondary goal, we also investigate the feasibility of incorporating hearing profiles and predicting individual hearing-impaired listeners' performance. 

In previous work~\cite{karbasi2015microscopic} we have presented a related, but simpler method of microscopic SI prediction with limited applicability. In this paper, we propose to improve the model using new objective measures. %Previously, we proposed to take the logarithm of the likelihood ratio between the true transcription and the best ASR word hypothesis as a stand-alone objective measure for SI prediction. For this purpose, we propose to use confidence measures, extracted from ASR, in combination with neural networks (NN), to predict the individual performance of normal-hearing and hearing-impaired subjects in listening experiments. 
For this purpose, we additionally include two ASR-based discriminance measures for predicting the speech intelligibility from a microscopic viewpoint~\cite{karbasi_CHAT17} and we improve their computation to arrive at a fully blind framework which only requires pre-trained reference models in its computation. Specifically, its low-level intelligibility measures are extracted utilizing an HMM-based ASR system. To create a full framework for SI prediction, we combine our new model-based intelligibility measures with a signal-based measure and add a neural-network-based regression stage. 
We finally demonstrate that the ultimately proposed \emph{NO-Reference Intelligibility} (\NORI{}) framework outperforms traditional methods and is broadly applicable, both under noisy conditions and for modeling the perception of hearing-impaired listeners.

The proposed measures and the full \NORI{} framework are explained in detail in the following section. 
The data used for evaluation are described in Section~\ref{sec:Databses}, the experimental setup and the evaluations will be presented and discussed in Sections~\ref{sec:experiment} and~\ref{sec:discus}.  

%We have previously proposed an objective measure for speech intelligibility prediction~\cite{karbasi2015microscopic}, namely the logarithm of likelihood ratio of the true transcription and the ASR-recognized word hypothesis given the input speech signal. Later, we have introduced two other heuristic measures~\cite{karbasi_CHAT17} for predicting the speech intelligibility from a microscopic point of view. The proposed measures are extracted utilizing an HMM-based ASR system. As an extension to this work, here, we propose to use confidence measures, extracted from ASR, in combination with neural networks (NN), to predict the individual performance of normal-hearing and hearing-impaired subjects in listening experiments. The proposed measures and the established framework are explained in detail in the next section. 
%============================================================================= 
\add{\section{Methods and Materials}}\label{sec:Method}
An automatic speech recognizer delivers, in principle, the same output measure, and it faces  the same challenges as a human listener when it comes to recognizing speech in a formalized intelligibility test.
%It is therefore of natural interest to investigate automatic speech recognition (ASR) as a non-intrusive prediction method for speech intelligibility. If the ASR framework models the individual human well, we expect a high correlation between human and machine hearing ability. Furthermore, in a second step we also can individualize the ASR to predict hearing-impaired \add{listeners' ability in SI tests}.   
In previous studies~\cite{FADE-schadler2015,FADE_ref,spille2018DNN} it has been suggested to use ASR as a predictor of  intelligibility and to compare the results directly to the results from human listeners. \delete{However, because of the discussed differences between human and machine hearing, the performance of an ASR often does not correlate well with the human listener under many circumstances.} 
\add{However, there are fundamental differences between human and machine hearing. 

Direct ASR-based predictions are restricted by the quality and the level of detail of the model of the human hearing system, and it is currently not possible to model human hearing accurately enough. Also, adding specific details of human perception does not necessarily contribute to better speech recognition performance: the recognition engine and language models are also different between ASR and humans listeners. Furthermore, ASRs are usually not trained to have the same performance (high or low) as human listeners, but they rather try to achieve the best performance overall. This leads to low correlations between the ASR recognition output and the human listener performance in many circumstances.}  
In order to overcome this problem, we propose to proceed differently in this paper: first, we will compute a set of ASR discriminance scores \add{(here called model-based measures)}---which are inspired by ASR confidence measures---\add{and compare them in microscopic SI prediction tasks in order to find the best-performing model-based measure. In addition, we also compute signal-dependent (or \emph{signal-based}) measures, including blindly-estimated SNR, STOI, ESTOI, mr-sEPSM, and the non-parametric estimation of MI using k-nearest neighbors (MI-KNN)~\cite{Taghia2014}. This allows for a detailed comparison between the traditional and proposed approaches. Moreover, a blind estimator of SNR can be included in our proposed non-intrusive SI prediction framework. This leads to a vector-valued measure that is a combination of the estimated SNR and the best-performing model-based measure.
In a second step, we combine each intelligibility measure with a subsequent prediction stage, which maps the respective measure to the expected human outcome in recognizing the speech units.}

\add{\subsection{Automatic speech recognition}} \label{subsec:ASRsetup}
At the first stage of this ASR system, speech-related features were extracted. Here, the first 13 Mel frequency cepstral coefficients (MFCCs), their first ($\Delta$) and second-order derivatives ($\Delta\Delta$) were used as features. Hamming-windowed frames with a length of 25 ms and a frame shift of 10 ms were used for the MFCC extraction algorithm. The sampling frequency was 25 kHz.

\add{We are using a conventional, statistical model-based automatic speech recognition (ASR) in all subsequent work. Similar to the procedure developed previously in our lab~\cite{karbasi_CHAT17},} each word $\nu$ was modeled using a linear left-to-right HMM.
% $\lambda_\nu$, resulting in 51 whole-word HMM models plus one silence model for the entire Grid corpus.
The number of states was chosen as three times the number of phonemes of the word. A 2-mixture diagonal covariance GMM was used to model the state output distribution of the HMMs.
In order to be able to compute confidence intervals,
% and to use the entire database during the intelligibility measure evaluation, 
each experiment was implemented with a 5-fold cross-validation \add{to train and test the ASR models} 5 times on different splits of the dataset. For each experiment and in each fold, the speech material was divided into disjoint training (60\%), development(20\%), and test sets (20\%). \add{To achieve the highest accuracy,} noise-dependent models were trained separately for each SNR. The development sets were only used to control the progress of the training. \add{In each fold, only the test set was used to extract the ASR-based intelligibility measures.}

\add{\subsection{Model-based SI measures}}\label{subsec:modelBasedMeasures}
There are many ASR-derived measures that can be used to estimate the human recognition performance. Based on the central idea that the degree of ASR uncertainty is related to the intelligibility level of the speech signal, we investigate several statistical measures: the time alignment difference (TAD) and the normalized likelihood difference (NLD)~\cite{karbasi_CHAT17}, the entropy (H), the log-likelihood ratio (L) and the dispersion (D). 

All of these measures are computed in parallel by the ASR system, which can provide several hypotheses with an associated likelihood as the recognized transcription of input speech.
%All measures are computed using those ratings as either the difference between the automatically recognized best hypothesis and the hypothesis matching the ground-truth transcription, or the difference between the best two consecutive hypotheses.
In contrast to the NLD and TAD~\cite{karbasi_CHAT17}, which require ground-truth time alignments and transcriptions of the signal, the newly proposed discriminative measures are computed based only on the information provided by the ASR and do not require any type of reference.
\add{Note that all measures in this paper are extracted at the word level. However, the proposed framework is not subject to any inherent constraints regarding the chosen speech units and it can also be generalized to a phoneme-based framework in future work.}

In the following subsections, the details of \delete{the implemented framework will be described.}
\add{the proposed method for extracting the model-based intelligibility measures will be described.}

\subsubsection{Preprocessing}\label{subsec:preproc}
\begin{figure}
	\centering
	\includegraphics{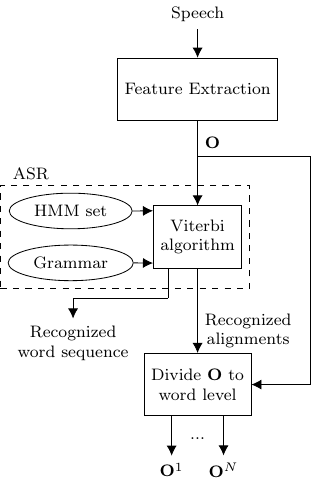}
	\caption{Block diagram of the preprocessing steps to produce observation sequences.} \label{fig:preProc}
\end{figure}

After applying feature extraction to the speech signal, the entire sequence of feature vectors, i.e. the \emph{observation sequence} $\mathbf{O}$, is divided into segments, each corresponding to one word. To perform this procedure, the word boundary information is required. This can be done in two ways: if the reference word alignment information is available, it can be employed directly to divide the observation sequence into words. If not, the automatic speech recognizer can be used to produce estimated word alignments through its recognition (see Fig.~\ref{fig:preProc}). In order to produce these recognized alignments, the trained HMM set and the grammar are used on the observation sequence in a Viterbi algorithm, yielding not only the recognized word sequence, but also the corresponding word boundaries. The Viterbi algorithm uses dynamic programming to find the HMM state sequence that matches the observation sequence best~\cite{viterbi}. 

After these steps, the intelligibility measures for each segment of the observation sequence $\mathbf{O}^n$ are calculated separately. In the following sections, the segment index $n$ is dropped for simpler notation, but note that all proposed measures are extracted given the segmented observation $\mathbf{O}^n$ at the word level. 

%%%%%%%%%%%%%%%%%%%%%%%%%%%%%%%%%%%%%%%%%%%%%%%%%%%%%%%%%%%%%%%%%%%%%%%%%%%%%%%%%%
%%%%%%%%%%%%%%%%%%%%%%%%%%%%%%%%%%%%%%%%%%%%%%%%%%%%%%%%%%%%%%%%%%%%%%%%%%%%%%%%%%

\add{\subsubsection{Extracting intelligibility measures}}\label{subsec:ExtractingMeasures}
We are considering 5 ASR-based measures, the dispersion, the entropy, the likelihood ratio, the time alignment difference and the normalized likelihood difference, which are introduced in detail below.

a) Model-based dispersion represents the degree of uncertainty of the ASR decoder in recognizing a speech signal. The dispersion of the speech signal corresponding to a single word is computed~\cite{dispersion} as 
\begin{equation}\label{eq:Disper1}
\textrm{D}=\frac{2}{N(N-1)}\sum_{k=1}^{N}\sum_{l=k+1}^{N}\log\frac{P\big(\lambda_{k}|\mathbf{O}\big)}{P\big(\lambda_{l}|\mathbf{O}\big)}.\\
\end{equation}
Here, $P\big(\lambda_k|\mathbf{O}\big)$ is the probability of the word model $\lambda_k$ given the observation sequence $\mathbf{O}$ where the probabilities are sorted in descending order, with $P(\lambda_{k=1})$ as the highest. $N$ is the number of best hypotheses used to compute the dispersion. 

In Eq.~\ref{eq:Disper1}, the model probabilities $P\big(\lambda_{k}|\mathbf{O}\big)$ are required. Having one HMM per word $k$, it is possible to compute the likelihood of the observation sequence given the word model, $P\big(\mathbf{O}|\lambda_k\big)$, using the forward algorithm~\cite{forward}, and to then obtain the model probability $P\big(\lambda_{k}|\mathbf{O}\big)$ using Bayes' theorem 
\begin{equation}
P\big(\lambda_{k}|\mathbf{O}\big)=\frac{P\big(\mathbf{O}|\lambda_k\big)P\big(\lambda_k\big)}{P\big(\mathbf{O}\big)}. \label{eq:Bayes}
\end{equation}
\add{The prior probability of the models $P\big(\lambda_k\big)$ can be acquired using a language model.}

\add{In our experiments we use a matrix test for evaluation.} Since the prior probability of each model $P\big(\lambda_\nu\big)$ is thus equal for all possible words and since the probability of the observation sequence $P\big(\mathbf{O}\big)$, is independent of $\lambda$, \add{in this work,} Eq.~(\ref{eq:Disper1}) can be reformulated to
\begin{equation}\label{eq:Disper2}
\textrm{D}=\frac{2}{N(N-1)}\sum_{k=1}^{N}\sum_{l=k+1}^{N}\log\frac{P\big(\mathbf{O}|\lambda_k\big)}{P\big(\mathbf{O}|\lambda_l\big)}.\\
\end{equation}

The steps required for extracting the HMM-based dispersion are shown in Algorithm~\ref{Disper_code} as an overview.
First, the likelihood of the observation $\mathbf{O}$ is needed for all word models $\lambda_\nu, \nu=1 \dots V$. These likelihoods need to be sorted in descending order, of which the $N$ highest values are used in Eq.~\ref{eq:Disper2} to compute the dispersion.

\begin{algorithm}[H] % enter the algorithm environment
	\caption{Compute the model-based dispersion for observation sequence $\mathbf{O}$, corresponding to the word position $n$ in a sentence.} % give the algorithm a caption
	\label{Disper_code} % and a label for \ref{} commands later in the document
%\setstretch{2}
	\begin{algorithmic}[1] % enter the algorithmic environment
		\State Compute the likelihood of $\mathbf{O}$, given all possible word models for the word position $n$, using the forward algorithm: $P\big(\mathbf{O}|\lambda_\nu\big), \nu = 1 ... V^n;$
		\State Sort all probabilities $P\big(\mathbf{O}^n|\lambda_\nu\big)$ in descending order; 
		\State Compute the dispersion for the $N$ highest probabilities, via Eq.~\ref{eq:Disper2};
		
	\end{algorithmic}
\end{algorithm}

b) The second measure proposed as a model-based intelligibility measure is the entropy, which can also be considered as an indicator of the ASR confidence. The entropy is computed for the segmented observation $\mathbf{O}$ via
\begin{equation}\label{eq:Entropy}
\textrm{H}=\sum_{m=1}^{M}-P\big(\lambda_{m}|\mathbf{O}\big)\log{P\big(\lambda_{m}|\mathbf{O}\big)},\\
\end{equation}
where $M$ is the number of all possible word models. 

c) The log-likelihood ratio $L$~\cite{karbasi2015microscopic} is the \delete{final} \add{third} proposed model-based measure. $L$ indicates the decoder's discrimination between the first and second best model. Therefore, $L$ corresponds to the dispersion measure where $N=2$.
For comparison, the results of intelligibility prediction using the $L$ measure are reported in Section \ref{sec:experiment}. 

\add{d) TAD and e) NLD are the final model-based measures. As introduced in~\cite{karbasi_CHAT17}, TAD is defined as the difference between the recognized and the ground-truth time alignments of the input signal. NLD is the normalized likelihood difference between the first and second most-likely models given the ground-truth transcription. Both measures require knowledge about the ground truth transcription and time alignments and can not be computed without a time-alignment and transcription reference.
}
%to compute the L via
%\begin{equation}\label{eq:L}
%\textrm{L}=\log\frac{P\big(\mathbf{O}|\lambda_{\nu^{\star(1)}}\big)}{P\big(\mathbf{O}|\lambda_{\nu^{\star(2)}}\big)}.\\
%\end{equation}
%Here, $\lambda_{\nu^{\star(1)}}$ and $\lambda_{\nu^{\star(2)}}$ represent the first and the second best word model, respectively.

\subsection{Signal-based SI measures}\label{subsec:signalBasedMeasures}   
In addition to the above model-based measures, the SNR is estimated as a signal-dependent measure (S$\widehat{\textrm{N}}$R). To estimate the SNR, the clean speech and noise signal powers are required. Here, the Improved Minima Controlled Recursive Averaging (IMCRA) algorithm and the Wiener filter are used to estimate the power of the clean signal, $\hat{\textrm{s}}(t)$, and the noise, $\hat{\textrm{n}}(t)$, without need for a reference. S$\widehat{\textrm{N}}$R is then calculated by
\begin{equation}\label{eq:SNR}
\textrm{S}\widehat{\textrm{N}}\textrm{R}=10\log\frac{\sum_{t}^{}\left(\hat{\textrm{s}}\left(t\right)\right)^2}{\sum_{t}^{}\left(\hat{\textrm{n}}\left(t\right)\right)^2}.\\
\end{equation}
As baseline intrusive measures, STOI\add{, ESTOI, MI-KNN, and mr-sEPSM} values are also extracted and used for comparison. \add{To compute these measures (except S$\widehat{\textrm{N}}$R), the reference alignments have been used to divide the signals into word units. We will investigate and compare their performance to find the strongest baseline for our setup. Note that none of these methods are explicitly designed to predict the performance of hearing-impaired listeners, but we will use the best-performing measure as the baseline in predicting the hearing impaired listeners as well.}

\subsection{Databases}\label{sec:Databses}
We used three speech intelligibility databases to evaluate the performance of the considered measures. These databases are all based on the speech signals from the Grid corpus~\cite{Grid} and we used associated intelligibility scores collected by listening tests with human subjects. The Grid corpus originally contains clean speech signals and their time alignments. In total, there are 34,000 clean speech signals in this corpus, collected from 34 English speakers at the University of Sheffield. Each Grid utterance is a semantically unpredictable 6-word sentence with a fixed grammar: Verb(4)-Color(4)-Preposition(4)-Letter(25)-Digit(10)-Adverb(4), where the numbers in parentheses represent the number of available choices for each word position.

%The first database (DB1) used here is a noisy version of original Grid, named Grid speech intelligibility database~\cite{Grid_Intelligibility}. It contains normal-hearing listeners' results in a keyword recognition listening test.
%Another noisy version of Grid speech signals, created with two other noise types and their correspondent intelligibility scores, is used here as the second database (DB2). The human intelligibility scores in this database have been previously collected by the authors~\cite{THMM-STOI} via crowdsourcing. The last database (DB3) contains hearing-impaired listeners' intelligibility scores in recognizing the noisy Grid speech signals from DB1.
%The structure of the Grid corpus and the databases DB1-3 are explained in detail in the next sections.

%\subsection{The Grid corpus} \label{subsec:Grid}
%The original Grid corpus contains 34000 clean speech signals in total, collected from 34 English speakers at the University of Sheffield. Each Grid utterance is a semantically unpredictable 6-word sentence with a fixed grammar: Verb(4)-Color(4)-Preposition(4)-Letter(25)-Digit(10)-Adverb(4), where the numbers in parentheses represent the number of available choices for each word position.  

\subsubsection{The \add{original} Grid intelligibility database (DB1)} \label{subsec:DB1}
The first database (DB1) is a noisy version of the original Grid corpus, the 'Grid speech intelligibility database'~\cite{Grid_Intelligibility}.
It contains noisy signals created by adding speech-shaped noise (SSN) to clean Grid signals at 12 different SNRs from -14\,dB up to 6\,dB with steps of 2\,dB, plus one condition at 40\,dB, labeled as 'clean'. Speech-shaped noise is created as Gaussian noise shaped with a long-term average spectrum identical to that of an averaged speech signal \add{from the Grid corpus}. DB1 also includes the listening test results from 20 normal-hearing listeners (NHL). The participants' responses to 2000 utterances and the ground-truth transcription are provided for the keywords color, letter, and digit.  

\subsubsection{The Grid intelligibility database with crowd-sourcing (DB2)}\label{subsec:crowdsoursingDB2}
A second noisy version of Grid speech signals was used in addition, referred to as DB2. The intelligibility scores in this database have been collected by the authors in previous work~\cite{THMM-STOI}.
In this database, in addition to speech-shaped noise (SSN), tokens with white noise and babble noise were also created, with the babble noise from the AURORA database~\cite{AURORA} containing a mixture of speech signals from a crowd of both female and male English speakers. 

To obtain human speech recognition results for the newly generated data, separate listening tests were carried out for the signals of each noise type in a large-scale listening experiment, using crowd-sourcing tests at~\citet{Crowdflower}.
Every test participant was asked to transcribe a set of 22 audio signals, containing different SNR conditions, ranging from -10\,dB to 6\,dB in steps of 2\,dB, where the noise type was fixed for a given test set. In order to prevent memory effects, we ensured that the same utterance was only utilized once within a given test set.

Experimentation in a weakly controlled environment like crowd-sourcing offers many benefits, specifically the potential large number of participants that can be approached, but this comes at the price of unknown variability in the sampling. In order to control quality of responses, therefore, additional control steps are necessary to ensure first that only participants are recruited who are qualified for the task (e.g., they need to have sufficient language skills) and to ensure that participants are concentrating during the tests. The English proficiency was established in a self-reporting questionnaire at the beginning, and to test concentration, each test set also contained 4 clean utterances. These were randomly interspersed between the actual test signals. For the analysis, only results were used where at least 50\% of the control utterances were correctly transcribed. Later analysis demonstrated that the minimum accuracy of listeners who passed this threshold, and were therefore used for further analysis, was above 70\%. \add{The hearing status of listeners in DB2 was not tested, because it is very difficult to measure or verify the listeners' hearing ability in crowd-sourcing experiments objectively.}

The transcriptions were recorded using a multiple-choice experiment using a web-based graphical user interface.
% shown in Fig.~\ref{fig:test_interface}.
%//////////////////////////////////////////  
%\begin{figure}
%	\centering
%%		\frame{\includegraphics[width=\linewidth]{pics/test_interface-eps-converted-to.pdf}}%natwidth=1631px,natheight=387px,
%	\frame{\includegraphics[width=1\columnwidth]{pics/test_interface-eps-converted-to.pdf}}
%	\caption{Test interface in the crowd-sourcing listening experiment.}
%	\label{fig:test_interface}
%\end{figure}
%//////////////////////////////////////////  
Each contributor was allowed to participate multiple times but restricted to 6 test sets. Participants were payed \$0.01 for each utterance. 

We collected the responses from 849 individual participants, who passed the quality control during the test. This corresponds to an overall number of 36,018 transcribed utterances in total.

\subsubsection{The Grid intelligibility database for hearing-impaired listeners (DB3)}\label{subsec:listeningtestDB3}
In order to evaluate the performance of the proposed measures in predicting the performance of hearing-impaired listeners (HIL), an additional intelligibility database was collected by conducting listening tests with hearing-impaired participants at the University of Southampton. The resulting third database (DB3) contains the same noisy Grid tokens as DB1, but with hearing-impaired listeners' intelligibility scores. 

All participants in this study were native English speakers and regular users of hearing aids. During the test, they were not wearing the hearing aid. First, pure tone audiometry (PTA) was performed to measure the hearing thresholds. The better ear was used in the speech test. In total, 9 listeners (3 female and 6 male) aged from 62 to 79 years took part in this study and were paid for their participation. The study was approved by the local ethics committee. 
Information of all participants including their individual audiometric thresholds are listed in Tab.~\ref{tab:PTA}.
%\add{The PTA experiments were performed in two separate time intervals. In the first half, the thresholds at 3 and 6 kHz were not measured. As a result, those frequencies are marked as not-available (n/a) in Tab.~\ref{tab:PTA} for the listeners who were measured in the first time interval.}
Speech from DB1 was presented, so  the stimuli contained SSN at SNRs ranging from -6\,dB to 6\,dB in steps of 2\,dB plus the clean signals. The stimuli were presented via circumaural headphones (Sennheiser HD380pro) in a quiet room. The equipment was calibrated with clean speech to a presentation level of 65 dB SPL using a sound level meter (Br{\"u}el\&Kj{\ae}r 2260) and artificial ear simulation (Br{\"u}el\&Kj{\ae}r 4153). 

The participants were asked to repeat the Grid keywords (color, letter and digit) after they had heard the whole sentence and the experimenter recorded their answers for each keyword. A short training session was performed prior to the main test in order to familiarize the participants with the task and the stimuli.
%-----------------------------------------------------------------------------
%\begin{figure}
%	\centering
%	\includegraphics[]{pics/meanPTAs.eps}
%	\caption{Average audiometric thresholds of the tested ears for all participants and their standard deviation depicted by the error bars.}
%	\label{fig:AveragePTAs}
%\end{figure}
%-----------------------------------------------------------------------------
%%-----------------------------------------------------------------------------
%\begin{figure}[htbp]
%	\centering
%	\includegraphics[width=0.45\textwidth]{pics/allPTAs.eps}
%	\caption{Audiometric thresholds of the tested ears for all participants.}
%	\label{fig:allPTAs}
%\end{figure}
%%-----------------------------------------------------------------------------
\begin{table*}[tb]
	%\vspace{-4mm}
	\caption{\label{tab:PTA}Audiometric thresholds of the tested ears for each listener.}
	%\vspace{3mm}
	\centerline{
%		\begin{tabular}{ L{1cm}  L{1cm}  L{1.5cm}  L{2cm}  C{1cm}  C{1cm}  C{1cm}  C{1cm}  C{1cm}  C{1cm}  C{1cm}  C{1.5cm} }
%			\hline
%			&     &        &            & \multicolumn{8}{ c }{ Thresholds (dB HL)}    \\
%			ID & Age & Gender & Tested ear &  0.25 & 0.5 & 1 & 2 & 3 & 4 & 6 & 8 (kHz)\\
%			\hline\hline
%			L1 & 73 & Male   & Right & 15 & 10 & 15 & 20 & n/a & 50 & n/a & 75 \\
%			L2 & 79 & Male   & Right & 35 & 30 & 40 & 50 & 50 & 65 & 70 & 70 \\
%			L3 & 66 & Female & Left  & 20 & 20 & 35 & 45 & n/a & 35 & n/a & 40 \\
%			L4 & 72 & Male   & Right & 0  & 5  & 15 & 45 & 60 & 65 & 55 & 60 \\
%			L5 & 62 & Male   & Left  & 10 & 15 & 20 & 25 & 40 & 50 & 55 & 60 \\
%			L6 & 65 & Female & Left  & 65 & 65 & 65 & 60 & 55 & 60 & 65 & 80 \\
%			L7 & 72 & Male	  & Left  & 15 & 30 & 40 & 50 & n/a & 60 & n/a & 65 \\
%			L8 & 72 & Male   & Right & 20 & 25 & 35 & 30 & n/a & 60 & n/a & 50 \\
%			L9 & 65 & Female & Right & 15 & 20 & 25 & 20 & n/a & 15 & n/a & 20 \\
%			\hline
%		\end{tabular}
		\begin{tabular}{ L{1cm}  L{1cm}  L{1.5cm}  L{2cm}  C{1cm}  C{1cm}  C{1cm}  C{1cm}  C{1cm}  C{1.8cm} }
			\hline
			&     &        &            & \multicolumn{6}{ c }{ Thresholds (dB HL)}    \\
			ID & Age & Gender & Tested ear &  0.25 & 0.5 & 1 & 2 & 4 & 8 (kHz)\\
			\hline\hline
			L1 & 73 & Male   & Right & 15 & 10 & 15 & 20 & 50 & 75 \\
			L2 & 79 & Male   & Right & 35 & 30 & 40 & 50 & 65 & 70 \\
			L3 & 66 & Female & Left  & 20 & 20 & 35 & 45 & 35 & 40 \\
			L4 & 72 & Male   & Right & 0  & 5  & 15 & 45 & 65 & 60 \\
			L5 & 62 & Male   & Left  & 10 & 15 & 20 & 25 & 50 & 60 \\
			L6 & 65 & Female & Left  & 65 & 65 & 65 & 60 & 60 & 80 \\
			L7 & 72 & Male	  & Left  & 15 & 30 & 40 & 50 & 60 & 65 \\
			L8 & 72 & Male   & Right & 20 & 25 & 35 & 30 & 60 & 50 \\
			L9 & 65 & Female & Right & 15 & 20 & 25 & 20 & 15 & 20 \\
			\hline
		\end{tabular}
	} 
	\vspace{1cm}
\end{table*}
%-----------------------------------------------------------------------------

\section{Experiments and results}\label{sec:experiment}
\add{Three sets of experiments were conducted to evaluate the performance of the proposed intelligibility measures for different tasks:
\begin{enumerate}[label=\Alph*.]
		\item Microscopic prediction of intelligibility for NHLs,
		\item Macroscopic prediction of intelligibility for NHLs,
		\item Microscopic prediction of intelligibility for HILs.
\end{enumerate} 
The overall goal of the experiments was to predict the performance of human listeners in keyword recognition tasks using ASR-driven measures and to analyze their performance based on both microscopic and macroscopic models.
For comparison, the intrusive signal-based measures STOI, ESTOI, MI-KNN and mr-sEPSM were also computed using the noisy speech and its clean counterpart. For every noisy speech utterance in each database considered in the current work, there is a clean counterpart from the original Grid corpus. Those pairs are used in the computation of intrusive measures.
} 

\add{\subsection{Microscopic SI prediction for normal-hearing listeners}} \label{subsec:NHL}
\add{For microscopic evaluation, the human listeners' performance was predicted using the estimated intelligibility measures as described in detail below and the outcomes are reported as 'accuracy' values, that is, the percentage of words for which the model predicts the human listener's performance correctly, either as ``recognized'' or as ``not recognized''.

In order to obtain the predicted intelligibility, a mapping was performed  between the speech intelligibility measure under test and the results of human speech recognition experiments, cf. Fig.~\ref{fig:mappingStage}. For that purpose, a binary classification neural network (NN) was trained to predict the human keyword recognition outcomes using the MATLAB patternnet toolbox. A feed-forward network was employed using the cross entropy as the cost function. The network had one hidden layer with 10 neurons. The input size was defined by the intelligibility measure(s) used for training and testing. In our experiments, it was either one scalar input, or two, in the case of \NORI. The NN was trained using the respective intelligibility measure for each token and its correspondent label, taken from the listening test results. The binary label was defined on the word level, depending on whether or not the word was correctly recognized by the human listener. A 7-fold cross validation was performed in this step so as to be able to compute confidence intervals for the evaluation results. At each fold, 70\% of data was used for training the network, 15\% for validation and 15\% for evaluation. The network was trained individually for each intelligibility measure, at each noise condition and over all SNRs.  

%%%%%%%%%%%%%%%%%%%%%%%%%%%%%%%%%%%%%%%%%%%%%%%%%%%%%%%%%%%%%%%%%%%%%%%%%%%%%%%%%%
\begin{figure}
	\centering
	\includegraphics{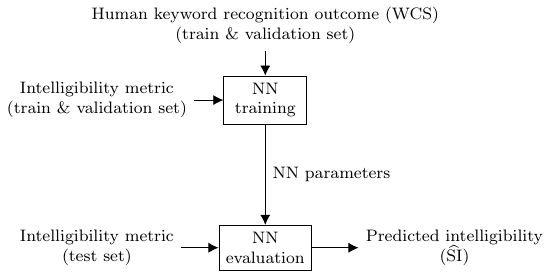}
	\caption{Block diagram of the mapping stage, including the training and testing phase to map the intelligibility measure to actual prediction values.} \label{fig:mappingStage}
\end{figure}
%%%%%%%%%%%%%%%%%%%%%%%%%%%%%%%%%%%%%%%%%%%%%%%%%%%%%%%%%%%%%%%%%%%%%%%%%%%%%%%%%%
}

\begin{figure}[htbp]
	\centering
	\includegraphics[width=0.75\columnwidth]{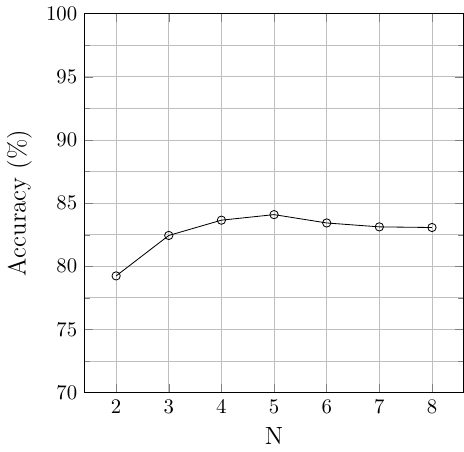}
	\caption{Accuracy of the measure dispersion, ($D$), for different values of $N$.}
	\label{fig:preExp1}
\end{figure}

\add{\subsubsection{Preliminary experiments}} \label{subsec:preExperiment}
\add{Prior to the main evaluation, two pilot experiments were conducted, using the data of DB1 only. In the first of theses, a pilot microscopic experiment was conducted in order to obtain the best value for $N$ in Eq.~\ref{eq:Disper1} was determined in a microscopic intelligibility prediction task, with $N$ varying between 2 and 8 (Fig.~\ref{fig:preExp1}). The best microscopic SI prediction accuracy for the dispersion was achieved at $N=5$ hypothesis, and $N=5$ was thus selected for further processes. This indicates that (at least for this dataset) the likelihood differences between the 5 best hypotheses contain the highest amount of information regarding the associated intelligibility, and adding more hypotheses does not provide further benefit.}
%%%%%%%%%%%%%%%%%%%%%%%%%%%%%%%%%%%%%%%%%%%%%%%%%%%%%%%%%%%%%%%%%%%%%%%%%%%%%%%%%%%
%%%%%%%%%%%%%%%%%%%%%%%%%%%%%%%%%%%%%%%%%%%%%%%%%%%%%%%%%%%%%%%%%%%%%%%%%%%%%%%%%%%

\add{In the second pilot experiment, the average accuracy of all proposed model-based intelligibility measures was computed to pre-select the most accurate model-based measure.} The speech signal was divided into word segments using \add{true alignments for the intrusive measures. For the non-intrusive measures, ASR-recognized alignments were used to divide the signal.}
%Unlike our previously proposed \add{model-based measure}~\cite{karbasi_CHAT17}, the entire procedure of intelligibility prediction can be performed blindly without requiring any reference signal or transcription alignment.
The results of the second pilot experiment are shown in Tab.~\ref{tab:ASR_measureComparision}.  
Among all investigated measures, the dispersion ($D$) shows the highest accuracy with 84.04\% \add{and 85.89\%, using the recognized and true alignments, respectively. Also, it can be seen that all measures perform better using the true alignments than using the recognized ones.} As \add{described above}, $D$ is computed using the five highest model likelihoods. In contrast, the entropy $H$ considers all possible model likelihoods in estimating the decoder uncertainty and can contain redundant information, leading to loss of accuracy. On the other hand, the log-likelihood ratio $L$ takes only the two highest model likelihoods into account and might therefore underestimate the amount of uncertainty in the ASR. 
\add{Also, the comparison between the results gained by using the true alignments versus the recognized alignments shows that using the correct time alignments has the biggest impact on the performance of TAD. This outcome was expected, since TAD is computed based on the time alignments, so the correctness of the time alignments is important to its performance. On the other hand, the dispersion $D$ is less influenced by the correctness of alignment information.}
%Since the dispersion takes more hypothesis information into account, it is more prone to the errors in detecting word boundaries in a speech signal that happen during the automatic speech recognition process.
%%-----------------------------------------------------------------------------
\begin{table*}[tb]
	%\vspace{-4mm}
	\caption{ Results of the 2nd pilot experiment, for the pre-selection of metrics. These show the average accuracy of the proposed model-based intelligibility measures \delete{$NLD$, $TAD$, $D$, $H$, $L$,} in predicting the performance of 20 normal-hearing listeners \add{from DB1} in the Grid keyword recognition \add{task}. \add{$NLD$= normalized likelihood difference, $TAD$= time alignment difference, $D$= dispersion, $H$= entropy, $L$= log-likelihood ratio}}
	%\vspace{3mm}
	\centerline{
		\begin{tabular}{ C{4cm} C{1.2cm}  C{1.2cm}   C{1.2cm}  C{1.2cm}  C{1.2cm} }
			\hline
			 & $NLD$ & $TAD$  & $D$   &   $H$    & $L$\\
			\hline\hline
			True alignment & 78.85 &	80.88 &	85.89 &	81.25 &	79.71\\
			Recognized alignment & 76.16 &	77.65 &	84.09 &	78.44 & 78.64 \\
			\hline
		\end{tabular}
	} \label{tab:ASR_measureComparision}
	%\vspace{-2mm}
\end{table*}
%%-----------------------------------------------------------------------------

As a result of this preliminary assessment, we chose the dispersion $D$ \add{computed with the recognized alignments} as the most appropriate reference-free model-based measure for the remainder of this paper, and in the following we compare its performance against signal-based measures in various test scenarios under different noise conditions. 
In total, \add{in addition to the baseline intrusive methods,} we investigated three \add{non-intrusive} intelligibility measures: \delete{STOI,} dispersion $D$, estimated SNR (S$\widehat{\textrm{N}}$R) and a combination of $D$ and S$\widehat{\textrm{N}}$R (the combination in \NORI). 

\add{\subsubsection{Microscopic SI prediction results}}
In this section, intelligibility measures \add{computed} from the normal-hearing listener databases DB1 and DB2 are used for evaluation. All results are averaged over all listeners of each database. 
The results are organized by the type of keyword in the sentence.
Tab.~\ref{tab:ssn_KWtype} shows the accuracy of all considered measures for DB1. 
The keywords in the Grid corpus are of three different types: colors, letters, and digits. The keyword types differ with respect of their degree of difficulty, mainly due to different perplexity (4 different choices for colors vs.~25 letters and 10 digits) and duration. We expected the highest intelligibility prediction accuracy for the color category, since it also contains the longest words, which helps in the decoding phase. The lowest intelligibility estimation accuracy was expected in the letter category due to its high perplexity and because the letters are relatively short, both contributing to a higher degree of difficulty.
Accuracy results in Tab.~\ref{tab:ssn_KWtype} show that the combined (reference-free) \NORI{} system delivers a slightly better performance than the reference-based STOI, specifically in the letter and the digit category. \add{However, it is not statistically different than the performance of STOI.}
On average, it can also be seen that the single measures, dispersion $D$ or S$\widehat{\textrm{N}}$R alone, perform less well.
%However, in the process of computing the STOI, the reference clean signal is used, while the S$\widehat{\textrm{N}}$R and $D$ have been computed blindly. The ASR-extracted alignments have been used to divide the signal to smaller segments and no reference is used in computing these two measures.
Apparently the information captured by $D$ and S$\widehat{\textrm{N}}$R that are combined in the \NORI{} framework \delete{is complementary}, is jointly improving the accuracy of intelligibility prediction - without the need of a clean reference. 

\add{Among the intrusive measures, mr-sEPSM had the lowest accuracy in predicting the performance of listeners in SSN. The extended version of STOI (ESTOI) and the mutual information-based measure (MI-KNN) did not achieve a higher accuracy than STOI in this microscopic evaluation.}
%\delete{A statistical significance analysis using Fisher's exact test~\cite{FishersExactTest_agresti1992} shows that the improvement of the \NORI{} framework over the STOI-based system is statistically significant at a level of $p<0.05$.}

%%-----------------------------------------------------------------------------
\begin{table*}[tb]
	%\vspace{-4mm}
	\caption{Average accuracy of all considered intelligibility measures in predicting the performance of 20 normal-hearing listeners \add{from DB1} in a keyword recognition task, categorized by keyword type.
		%The dispersion $D$ has been computed using the automatically recognized alignments.
	}
	%\vspace{3mm}
	\centerline{
		\begin{tabular}{ C{2.5cm}   C{2cm} C{1.2cm} C{1.3cm} C{1.7cm} C{1.5cm} C{1.2cm}  C{1.5cm} }
			\hline
			Keyword type & mr-sEPSM & STOI  & ESTOI & MI-KNN   & $ D $      & S$\widehat{\textrm{N}}$R  &   \NORI{}\\
			\hline\hline 
			Color   	 &  86.92   & 88.84 &  88.21 &  87.18	&  88.53	&  87.22 &  88.67\\
			Letter   	 &  66.85   & 82.72 &  80.81 &  77.25	&  79.25	&  81.52 &  83.02\\
			Digit   	 &  77.75   & 85.34 &  83.35 &  79.21	&  84.49	&  84.44 &  85.75\\
			\hline                                                                            
			Average      &  77.17   & 85.63 &  84.12 &  81.21	&  84.09	&  84.39 &  \textbf{85.81}\\
			\hline
		\end{tabular}
	} \label{tab:ssn_KWtype}
	%\vspace{-2mm}
\end{table*}
%%-----------------------------------------------------------------------------

\begin{figure}[htbp]
	\centering
	\includegraphics[width=1\columnwidth]{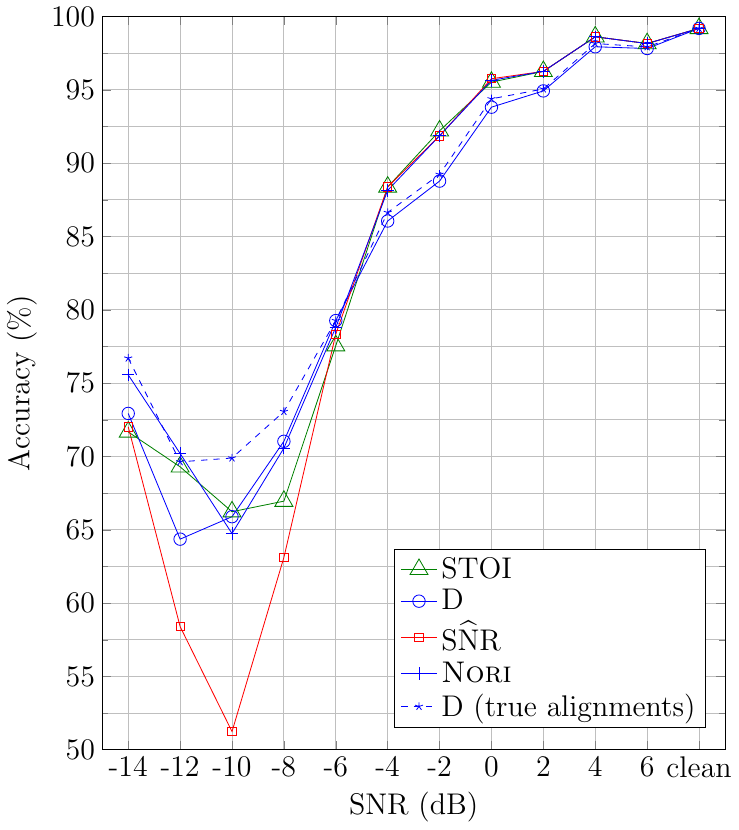}
	\caption{Accuracy of all considered intelligibility measures in predicting the keyword recognition performance of all listeners in DB1 with SSN data, using a feed-forward NN in the mapping stage.}
	\label{fig:ssn_SNR}
\end{figure}

The intelligibility prediction performance of the methods is assessed by computing their accuracy in predicting listening test results for different SNRs. \add{The results for STOI, as the best intrusive method, and for all non-intrusive methods} are shown in Fig.~\ref{fig:ssn_SNR}. All considered measures perform worse at around -10 dB SNR. This corresponds roughly to the human SRT, which is at -10.31 dB in DB1. This implies that (at least using our methods) predicting the intelligibility of a speech signal is most difficult at SNRs where human performance is around 50\%. Accuracy rises with higher SNRs as expected, but also rises for lower SNRs.
\add{STOI is performing slightly better than D in almost all SNRs, except for three low SNRs.}For comparison, \add{in addition to D, which is computed using the recognized alignments}, we also include a dispersion measure with true alignment. It shows better performance at lower \add{SNRs, but the improvement is very small at higher SNRs. This proves the importance of accurate word boundaries in computing the dispersion measure. The dispersion computed with true alignment is still performing slightly worse than STOI at higher SNRs.} In SNRs from -6 down to -14 dB, S$\widehat{\textrm{N}}$R shows lower accuracy than $D$. However, at higher SNRs it performs slightly better. The \NORI{} framework, taking advantage of both measures---$D$ and S$\widehat{\textrm{N}}$R---is performing well in all SNRs.

%%%%%%%%%%%%%%%%%%%%%%%%%%%%%%%%%%%%%%%%%%%%%%%%%%%%%%%%%%%%%%%%%%%%%%%%%%%%%%%%%%%
%%%%%%%%%%%%%%%%%%%%%%%%%%%%%%%%%%%%%%%%%%%%%%%%%%%%%%%%%%%%%%%%%%%%%%%%%%%%%%%%%%%%

\add{The above experiment was based on data from DB1, which contains only one noise type (speech-shaped noise) at 12 different SNRs. Incorporating more types of noise in speech distortion is important for a more comprehensive evaluation of the proposed \NORI{} method, since different types of distortion can influence the accuracy of intelligibility prediction methods differently.

Therefore, in the next experiment, the database DB2 was used for further investigation of the proposed measures. DB2 consists of speech signals distorted with three different noise types: speech-shaped noise, white noise and babble noise, with listening tests collected by crowd-sourcing, as detailed in Sec.~\ref{subsec:crowdsoursingDB2}.  
The average accuracy of the intelligibility measures is shown in Fig.~\ref{fig:allNoiseTypesDB2}.
Here, the dispersion measure $D$, solely and also in combination with S$\widehat{\textrm{N}}$R, outperforms the STOI measure in all noise conditions, especially in SSN and white noise. The combined \NORI{} measure shows the highest accuracy prediction among all tested measures.
Predicting intelligibility in white noise is the most difficult task; all white noise results show lower accuracy in comparison to babble and speech shaped noise. 
According to Fisher's exact test, the accuracy gained by using either $D$, or $D$+S$\widehat{\textrm{N}}$R in the full \NORI{} framework, is statistically significant over the use of the STOI measure at a level of $p<0.01$ in conditions with SSN or white noise. However, in babble noise, the accuracy achieved with the proposed measures is not statistically different from that based on the STOI measure.

Among all considered intrusive methods, STOI showed the best performance on DB1. In the next experiment we therefore only evaluated and compared our proposed measures against STOI as the best-performing intrusive measure.
%%%%%%%%%%%%%%%%%%%%%%%%%%%%%%%%%%%%%%%%%%%%%%%%%%%%%%%%%%%%%%%%%%%%%%%%%%%%%%%%%%%
\begin{figure}[t]
	\centering
	\includegraphics[width=1\columnwidth]{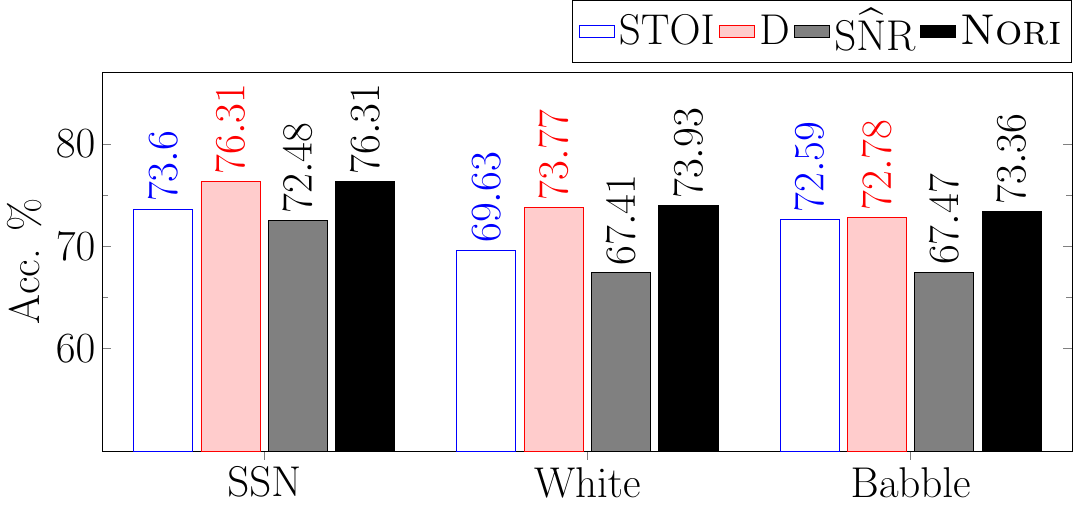}
	\caption{Average accuracy of all considered intelligibility measures in three different noise types, SSN, white, and babble noise, in predicting the performance of \delete{normal-hearing} all listeners in DB2.}
	\label{fig:allNoiseTypesDB2}
\end{figure}
%%%%%%%%%%%%%%%%%%%%%%%%%%%%%%%%%%%%%%%%%%%%%%%%%%%%%%%%%%%%%%%%%%%%%%%%%%%%%%%%%%%
}

\subsection{Macroscopic SI prediction for normal-hearing listeners}
In addition to the microscopic evaluation, we also analyzed the correspondence between the predicted and the ground truth intelligibility, i.e., the average human word recognition scores.
	
\subsubsection{Experimental setup}
As the intelligibility measures are extracted per keyword here, their corresponding ground truth are binary data. However, for macroscopic evaluation, a continuous distribution of the intelligibility scores is required. In our databases, each file contains one Grid utterance and each utterance contains 3 keywords. Before evaluation, the speech files were randomly divided into segments consisting of 10 files. Then, all measures, extracted per keyword in each segment, were averaged. This was repeated for every SI measure and the same averaging process was also applied to the corresponding human word recognition scores. This created continuously-distributed data for macroscopic evaluation. 
	 
As evaluation metrics, the averaged normalized cross-correlation coefficient (NCC), Kendall's Tau ($\tau$), and the root mean square error (RMSE) were used to compare the predicted intelligibility and human word recognition scores across different SNRs. NCC and RMSE only provide valid estimates when their input variables have a linear relationship. However, it is possible to linearize the relationship between the machine-derived and the human listening test results by estimating a mapping function. To estimate such a function, both neural networks and logistic function estimation were tested. Since the logistic function had a lower accuracy than the neural network regression, only the NN results are reported here. For NN training and testing, fitnet, the MATLAB shallow neural network toolbox, was used with the mean square error as its default cost function. A mapping function was estimated for each intelligibility measure using a network with one hidden layer with 10 neurons. For every measure and each noise type, a separate mapping was estimated over all SNRs. Since utilizing a mapping function can influence the final evaluation results, a metric that does not require linearity, namely Kendall's Tau, was also included in the evaluation. Kendall's Tau is computed between the rank ordering within two data sets without requiring a mapping function.
Similar to the microscopic experiments, the network parameter estimation and evaluation was performed in a 7-fold cross validation. This allowed us to use all available data for evaluation and achieve more reliable results. The results reported here are the average values of evaluation metrics computed over the ones obtained for each fold. 

This evaluation was performed with normal-hearing listener data from DB1 and DB2.

\subsubsection{Macroscopic SI prediction results}
In the following experiments, our intelligibility prediction, based on all considered SI measures, is being compared to the human speech recognition accuracy, given as the word correct score (WCS). The WCS is computed by dividing the number of correctly recognized keywords by the total number of keywords.

\begin{figure}[htbp]
	\centering
	\includegraphics[width=1\columnwidth]{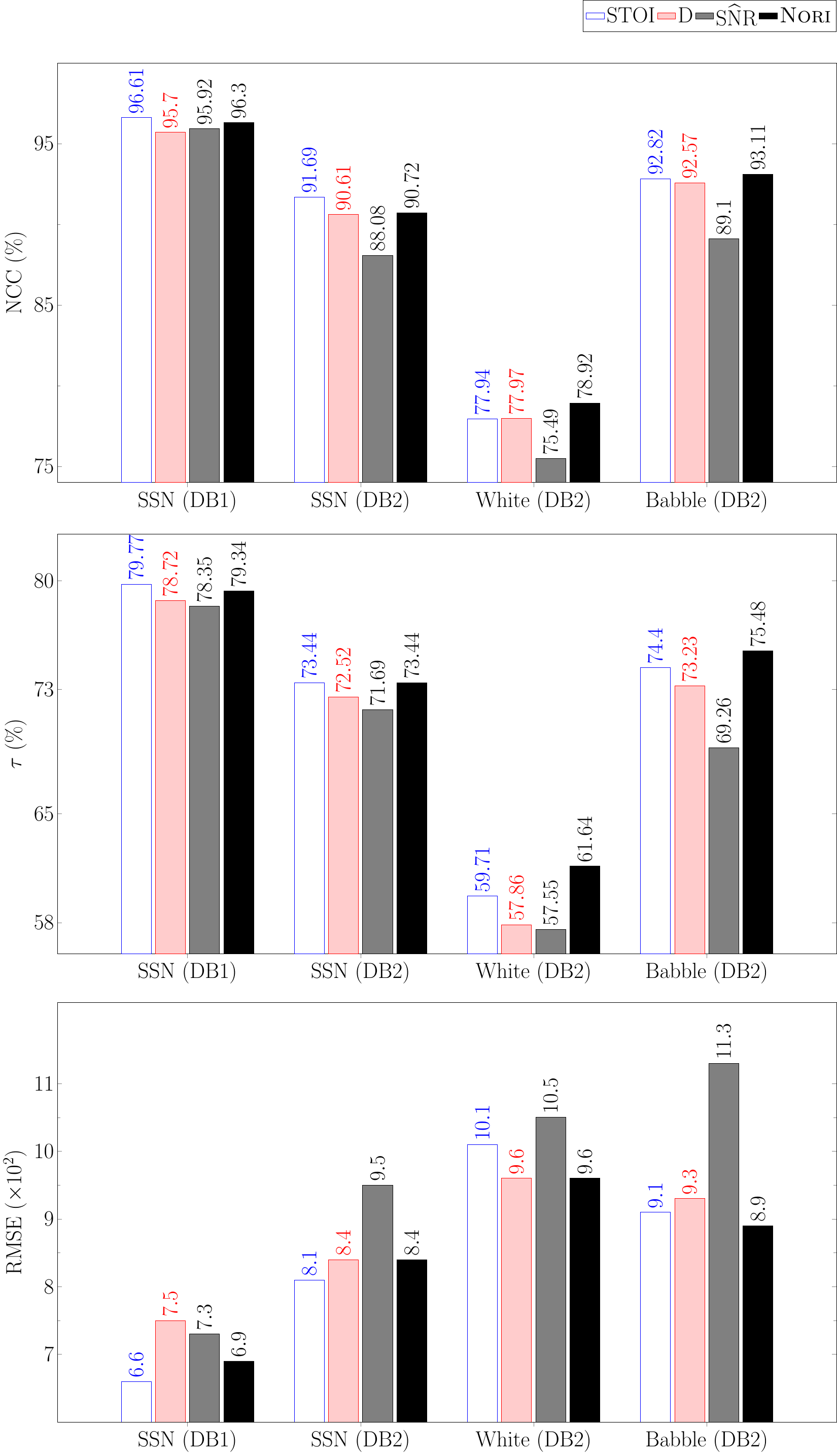}
	\caption{Comparison of intelligibility measures regarding their predictive performance for listening tests. Noise types are SSN, babble, and white noise \add{from DB2} in terms of NCC (\%), RMSE, and Kendall's Tau ($\tau$).}
	\label{fig:macroExperiment}
\end{figure}

The amount of correlation between the intelligibility prediction and the correspondent WCS (ground truth intelligibility) \add{in all conditions taken from DB1 and DB2} is shown in Fig.~\ref{fig:macroExperiment} in terms of NCC, $\tau$, and the root mean square error (RMSE).
The average performance evaluation shows that $D$+S$\widehat{\textrm{N}}$R, as in the \NORI{} framework, again performs better than the individual measures $D$ and  S$\widehat{\textrm{N}}$R in terms of correlation and error.
%%-----------------------------------------------------------------------------
\begin{table}[b]
	%\vspace{-4mm}
\delete{
	\caption{Average performance of all considered intelligibility measures in terms of NCC (\%), Kendall's Tau ($\tau$ \%), and RMSE, computed between the intelligibility measure predictions and listening test results (WCS) from DB1 with SSN data.}
	%\vspace{3mm}
	\centerline{
		\begin{tabular}{ C{2.5cm}   C{1.2cm}  C{1.2cm}  C{1.2cm}  C{1.8cm} }
			\hline
			performance measure & STOI    & $D$    &  S$\widehat{\textrm{N}}$R   &  \NORI{} \\
			\hline\hline
			NCC (\%)   	& \textbf{96.61}  & 95.70   & 95.92 & 96.30  \\
			$\tau$ (\%) & \textbf{79.77}  & 78.72   & 78.35 & 79.34  \\
			RMSE        & 0.066  & \textbf{0.075}   & 0.073 & 0.069  \\
			\hline
		\end{tabular}
	} \label{tab:SSN_corr} 
	%\vspace{-2mm}
}
\end{table}
%%-----------------------------------------------------------------------------

The average performance of all tested intelligibility measures was also evaluated, differentiating the three groups of noisy data present in DB2, shown in Fig.~\ref{fig:macroExperiment}.
In the case of white and babble noise, \NORI{} always outperforms STOI in terms of all three evaluation metrics. For SSN, \NORI{} performs slightly worse than STOI, however, with the added benefit of predicting the intelligibility non-intrusively. Similar results were achieved using the SSN data from DB1.
S$\widehat{\textrm{N}}$R has the lowest performance under all noise conditions.
%%%%%%%%%%%%%%%%%%%%%%%%%%%%%%%%%%%%%%%%%%%%%%%%%%%%%%%%%%%%%%%%%%%%%%%%%%%%%%%%%%%
%%%%%%%%%%%%%%%%%%%%%%%%%%%%%%%%%%%%%%%%%%%%%%%%%%%%%%%%%%%%%%%%%%%%%%%%%%%%%%%%%%%

%%%%%%%%%%%%%%%%%%%%%%%%%%%%%%%%%%%%%%%%%%%%%%%%%%%%%%%%%%%%%%%%%%%%%%%%%%%%%%%%%%%
\begin{figure}[htbp]
	\centering
	\includegraphics[width=1\columnwidth]{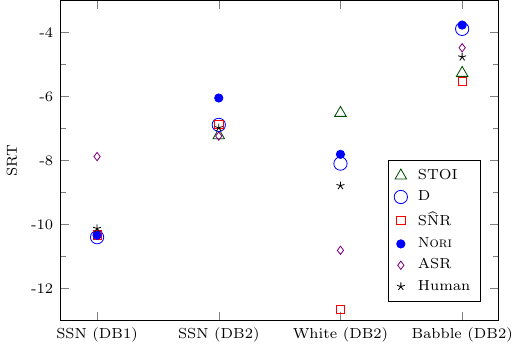}
	\caption{Predicted SRTs using all considered intelligibility measures, also introducing ASR recognition results, in comparison to the ground truth SRT computed from the human listening test results \delete{collected via crowdsourcing in DB2}. All SRT values have been reported for \delete{three} \add{four} groups of speech data distorted by \add{SSN in DB1,} SSN \add{in DB2}, white noise \add{in DB2,} and babble noise \add{in DB2}.}
	\label{fig:SRT_DB1DB2}
\end{figure}
%%%%%%%%%%%%%%%%%%%%%%%%%%%%%%%%%%%%%%%%%%%%%%%%%%%%%%%%%%%%%%%%%%%%%%%%%%%%%%%%%%%

The speech reception threshold (SRT) -- the SNR at which the speech recognition accuracy is 50\% -- is \delete{a succinct} \add{another} measure to describe the intelligibility of a signal. \add{It is computed based on the psychometric function over many different SNRs but it does not provide detailed information about the intelligibility of the signal at each SNR separately.}
Fig.~\ref{fig:SRT_DB1DB2} shows the SRTs computed using the \add{listening test} data collected \delete{with crowd-sourcing} in \add{DB1 and} DB2. \add{The estimated SRTs in SSN show large differences between DB1 (-10.1~dB) and DB2 (-6.9~dB). The listening tests in DB1 have been conducted in a controlled environment with normal-hearing listeners, while such a level of control was not possible in the crowd-sourcing experiment. Therefore, some differences are expected.} \add{The human results from DB2 show the} highest SRT in babble noise (at -4.8~dB), and the lowest in white noise (-8.8~dB). \add{As expected, babble noise is the most effective masker and white noise is the least effective masker for human listeners. On the other hand, predicting the human SRT is most difficult in white noise compared to babble and SSN.}
%The predicted values from all considered intelligibility measures and using the ASR recognition outputs are also shown in Fig. ~\ref{fig:SRT_DB2}. 
Comparing the predicted SRTs in each noise type shows that using the dispersion measure $D$ yields results closest to human performance in conditions with SSN and white noise, where the ASR-based prediction and STOI do not perform as well. For babble noise, the predicted SRT of $D$ is slightly less accurate than that based on the STOI and ASR values. Also, the results show that adding the estimated S$\widehat{\textrm{N}}$R to $D$ does not improve prediction accuracy in this case.
\add{Based on the results from SSN (DB1), it can be seen that using ASR as a direct predictor overestimates the SRT and performs less accurately than the other intelligibility measures in this case.}

\add{\subsection{Microscopic SI prediction for hearing-impaired listeners}} \label{subsec:HIL}
In the final set of experiments, we investigated how well we can predict whether an individual hearing-impaired listener will be able to understand specific utterances or words. 

\begin{figure}[htbp]
	\centering
	\includegraphics[width=1\columnwidth]{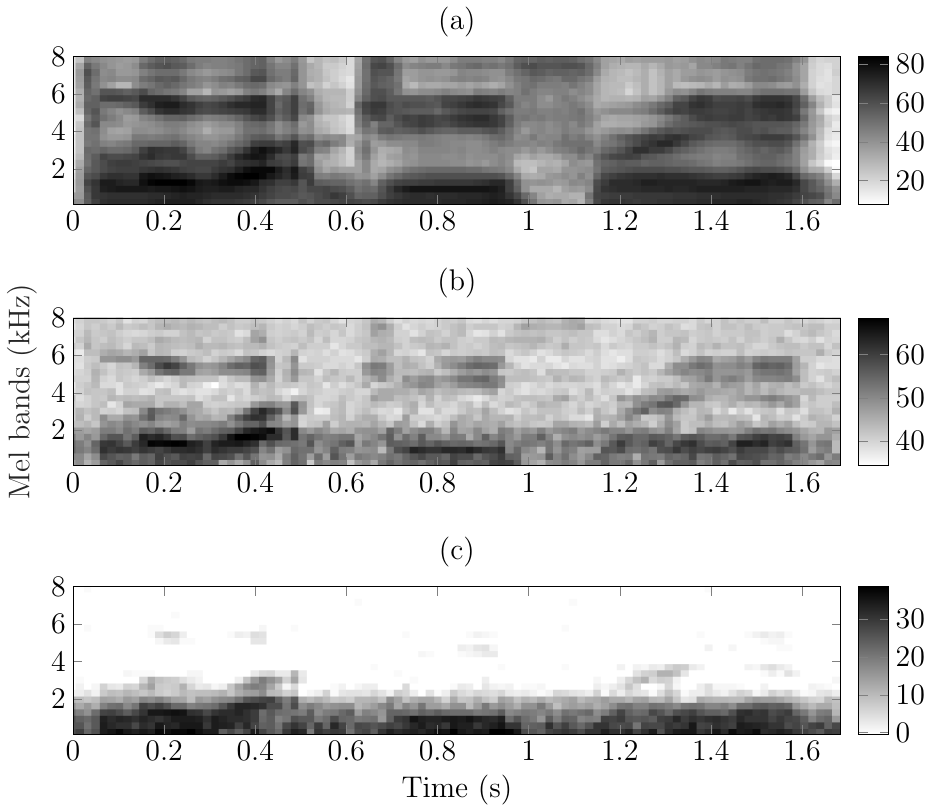}
	\caption{Short-time Mel-scale spectral representation of one speech signal in different scenarios; (a) clean , (b) noisy, and (c) threshold-adapted noisy spectrum.}
	\label{fig:specsGray}
\end{figure}

In order to simulate individual hearing status, personalized features were extracted~\cite{FADE_modelingHI_ref} \add{for the ASR-based SI measures}. To do this, the audiogram data was used to set thresholds in computing the logarithmic Mel-scale spectrograms during the MFCC feature extraction, prior to which, all speech signals in DB3 were amplified to the same hearing level of presentation. An example of these personalized feature maps is shown in Fig.~\ref{fig:specsGray}. Shown is the the Mel scale spectrogram of (a) a clean speech signal, (b) its equivalent representation at 4dB SNR, and (c) the threshold-adapted version. It can be seen how raised hearing thresholds cause \delete{smearing of the spectrum} \add{loss of information}, specifically at higher frequencies.  
	
\add{\subsubsection{Experimental setup}}
\add{
The outcomes are reported as 'accuracy' values, that is, the percentage of words where the model predicts the HIL performance on a single word correctly. Similar to the microscopic experiments on normal-hearing listeners, a NN-based mapping was applied to the SI measures before computing the accuracies. In this experiment, for each listener and noise condition, a separate mapping NN was trained over all SNRs.} 
\delete{For this purpose,} \add{For the model-based measures,} each listener was modeled separately, i.e., all HMMs and GMMs were trained for the specific listener. \add{The signal-based measures, STOI and S$\widehat{\textrm{N}}$R, however, were computed without further processing to model the hearing loss.}

\add{\subsubsection{Results for HI listeners}}
Intelligibility prediction results using STOI, $D$, S$\widehat{\textrm{N}}$R, and $D$+S$\widehat{\textrm{N}}$R as proposed in the \NORI{} framework, are shown in Tab.~\ref{tab:HI}. The results are reported individually and also on average. They show that \NORI{} outperforms STOI on average and also individually for participants L2, L6 and L9. For other HILs, \NORI{} performs almost at the same level as STOI.

\add{In evaluations of the FADE framework~\cite{FADE-schadler2015}, it has been shown that the audiogram data is not sufficient for modeling the hearing impairment effects on speech recognition results. In our experiments, however,} they have shown significant benefit for accurately predicting the speech perception of HILs, even without employing a reference signal, even despite the fact that our simple hearing impairment model only applies raised thresholds and ignores all other potential problems of the auditory system like widening filters or reduced temporal gap detection. With access to further supra-threshold hearing loss information, the feature extraction algorithm could, and should be adapted to achieve an even better prediction of individual listener performance.

%-----------------------------------------------------------------------------
%-----------------------------------------------------------------------------

%-----------------------------------------------------------------------------
\begin{table}[htbp]
	%\vspace{-4mm}
	\caption{\label{tab:HI} Average accuracy in predicting the hearing-impaired listeners' speech recognition performance.}
	%\vspace{3mm}
	\centerline{
		\begin{tabular}{ C{2cm}   C{1.2cm}  C{1.2cm}  C{1.2cm}  C{1.8cm} }
			\hline
			HIL ID & STOI    & $D$     & S$\widehat{\textrm{N}}$R &  \NORI{} \\
			\hline\hline
			L1 & 80.83 & 80.41 & 77.91 & 79.16\\
			L2 & 70.00 & 66.04 & 66.04 & 73.12\\
			L3 & 74.79 & 73.95 & 70.83 & 74.16\\
			L4 & 74.58 & 76.25 & 68.12 & 74.58\\
			L5 & 83.33 & 82.70 & 78.54 & 81.66\\
			L6 & 65.35 & 70.53 & 61.42 & 69.82\\
			L7 & 70.53 & 68.92 & 60.35 & 70.35\\
			L8 & 77.50 & 72.32 & 72.85 & 74.82\\
			L9 & 75.17 & 77.85 & 75.35 & 76.96\\
			\hline
			Mean & 74.67 & 74.33 & 70.15 & 74.95 \\
			\hline
		\end{tabular}
	} 
	%\vspace{-2mm}
\end{table}
%-----------------------------------------------------------------------------

%=============================================================================
\section{Discussion}\label{sec:discus}
We have introduced a novel reference-free approach to speech intelligibility prediction, and we have evaluated its performance for both normal-hearing and hearing-impaired listeners in various noise conditions. 

Our approach bases on an intelligibility measure that we derived from the discriminability (or confidence) information within an automatic speech recognition (ASR) system, namely the model-based dispersion $D$. Given that speech intelligibility is affected by many internal and external factors, we also took signal-based intelligibility information into account. It has been shown in previous studies that the SNR is one such indicator of speech intelligibility. Accordingly, we combined the discriminability score $D$ with the estimated S$\widehat{\textrm{N}}$R to provide an improved measure. 

We showed that the resulting \add{non-intrusive} \NORI{} method performs well in \add{predicting the performance of normal-hearing and hearing-impaired listeners in} a noisy keyword recognition task \delete{for normal-hearing and hearing-impaired listeners} and that \NORI{} can even outperform the often-used reference-based STOI measure. 

The results of evaluation with SSN data from DB1 (Fig.~\ref{fig:ssn_SNR}) show that this approach increased prediction accuracy in most SNRs. The average performance of all considered measures are statistically different from each other at a level of $p<0.05$ (Table ~\ref{tab:ssn_KWtype}) with \NORI{} outperforming STOI in two out of three conditions. 
As shown in Fig.~\ref{fig:ssn_SNR}, predicting speech intelligibility is most difficult when the word recognition rate of listeners is around 50\%, i.e.,~around the speech reception threshold (SRT), where the slope of the psychometric function is steepest. 
Our initial experiments demonstrate that if we provide the proposed method with additional reference alignments to segment the speech to smaller units, the ASR-based discriminance score $D$ becomes more accurate and outperforms STOI, even without taking S$\widehat{\textrm{N}}$R into consideration. Using the reference time alignments for speech segmentation also helps the dispersion to be more accurate in lower SNRs close to the SRT. This implies an expected benefit of further robustness improvements in the ASR system, which will be one goal of future work. 

The \NORI{} framework has been developed to predict the intelligibility of speech in smaller units like words. Among the considered keywords, letters and digits are the shortest words with the highest perplexity, which makes them the most difficult for ASR and also for other instrumental measures to predict their intelligibility. \delete{A categorized} \add{An} analysis based on the word type shows that the proposed  \NORI{} framework is more successful at word-level intelligibility prediction in comparison to STOI in these most difficult cases of letters and digits (Tab.~\ref{tab:ssn_KWtype}). 

We also investigated the ability of \NORI{} to predict individual hearing-impaired listeners' performance in various noise situations. Although our hearing-impairment model is very simple and only involves raised thresholds, \NORI{} performs equivalently to STOI in some conditions, and is even slightly superior on average. The prediction accuracy may be improved further when including other supra-threshold effects characterizing hearing impairment, like wider auditory filters and temporal smearing. \add{Note, however, that STOI, in contrast to \NORI{}, has not been explicitly designed for modeling hearing impaired listening.} 

Overall, the proposed intelligibility estimate of the \NORI{} framework is computed non-intrusively and is successful in predicting the speech intelligibility microscopically as well as macroscopically. Whilst the well-known STOI measure requires a clean reference signal, our framework only requires some time and data for training the ASR models. 

The proposed method has been evaluated on the Grid corpus, a small-vocabulary dataset with a matrix sentence test structure, and it is hence applicable directly to other similar data. However, the framework is not limited to matrix tests, but rather it can and should be extended and used for the prediction of intelligibility at a phoneme level in future work. Consequently, the framework is also extendable to large-vocabulary scenarios. This application, however, would call for a large-vocabulary speech dataset with a corresponding, large set of human listening test results collected for evaluation.

% * <bleeck@gmail.com> 2018-08-30T14:39:35.130Z:
%
% ^.
%=============================================================================
\section{Conclusion}\label{sec:Conc}
The main goal of this work was to predict microscopic (i.e. word-by-word) speech intelligibility in a non-intrusive manner, i.e., without access to any clean reference signal.
The dispersion, extracted as a discriminance measure from ASR models, together with the blindly estimated S$\widehat{\textrm{N}}$R were introduced as non-intrusive measures and embedded as predictors into the introduced \NORI{} framework for \emph{NO-Reference Intelligibility} estimation. 

The evaluation is based on a large number of normal-hearing listeners' data, showing that the \NORI{} framework can outperform the STOI in predicting NHL's word recognition performance and it correlates well with human data in terms of NCC, $\tau$ and RMSE in most conditions.
Overall, \NORI{} performs accurately and can predict the word-level speech intelligibility precisely, without the need for any extra information like the clean reference signal during the prediction process.

Finally, it was shown that it is feasible to use the proposed measure in predicting the performance of hearing-impaired listeners, reaching an accuracy equivalent to that of STOI. However, to better model the hearing impairments, it will be a goal of future work to take supra-threshold effects into account in computing the font-end features for the employed ASR models. To evaluate the framework with large-vocabulary data, it will be necessary to collect larger databases of real-life speech that are annotated with human listening results---a task that, while difficult, is deemed by the authors to be vital to move the field forward towards real-life on-line intelligibility optimization.

Research in speech communication systems requires the ability to assess speech intelligibility rapidly. Experiments with human listeners are time-consuming and costly, and they are unrealistic for big-data and learning-driven approaches and for on-line adaptation. Therefore, estimated speech intelligibility needs to be available and reliable as a stand-in measure. We have demonstrated here that reference-free algorithms are a viable option for this purpose, which can considerably widen the space for system optimization towards real-life and user-adaptive speech enhancement. 

%-------------------------------------------------
%Normal journal cite: \citep{joursamp1},
%% before appendix (optional) and bibliography:
\begin{acknowledgments}
This research has received funding from the European Union's Seventh Framework Programme FP7/2007-2013/ under REA grant agreement n$^\circ$[317521].
The authors would like to thank Jon Barker for providing a noisy version of the Grid database with comprehensive listening test results.
\end{acknowledgments}

% -------------------------------------------------------------------------------------------------------------------
%   Appendix  (optional)

%\appendix
%\section{Appendix title}

%If only one appendix, please use
%\appendix*
%\section{Appendix title}

%=======================================================
%IMPORTANT
\bibliography{Refs}
%\bibliography{sampbib}

%Use \bibliography{<name of your .bib file>}+
%to make your bibliography with BibTeX. 

%Once you have used BibTeX you
%should open the resulting .bbl file and cut and paste the entire contents 
%into the end of your article.
%=======================================================

\end{document}